\documentclass[12pt,preprint]{aastex}

\usepackage{graphicx}
\usepackage{epsfig,amsmath}

\newcommand{\be}{\begin{eqnarray}}
\newcommand{\ee}{\end{eqnarray}}

\begin{document}

%\title{Baryon Anisotropy within Gamma-Ray Burst Jets}
\title{The Case for Anisotropic Afterglow Efficiency within Gamma-Ray
Burst Jets}

\author{David Eichler\altaffilmark{1} and Jonathan Granot\altaffilmark{2}}
\altaffiltext{1}{Physics Department, Ben-Gurion University,
Be'er-Sheva 84105, Israel; eichler@bgu.ac.il}
\altaffiltext{2}{Kavli Institute for Particle Astrophysics and
Cosmology, Stanford University, P.O. Box 20450, Mail Stop 29,
Stanford, CA 94309; granot@slac.stanford.edu}

\begin{abstract}

Early X-ray afterglows recently detected by {\it Swift} frequently
show a phase of very shallow flux decay lasting from $\sim
10^{2.5}\;$s up to $\sim 10^4\;$s, followed by a steeper, more
familiar decay. We suggest that the flat early part of the light
curve may be a combination of the decaying tail of the prompt
emission and the delayed onset of the afterglow emission observed
from viewing angles slightly outside the edge of the region within
the jet with prominent afterglow emission, as predicted
previously. This would imply that a significant fraction of
viewers get very little external shock energy along their line of
sight, and, therefore, see a very high $\gamma$-ray to kinetic
energy ratio at early times. The early flat phase in the afterglow
light curve implies, in a rather robust and model independent
manner, a very large $\gamma$-ray efficiency, typically $\gtrsim
90\%$, which is very difficult to extract from baryons by internal
shocks.

\end{abstract}


\section{Introduction}

Although early models of fireballs \citep{Goodman86} did not postulate
baryons within, the existence of baryons in $\gamma $-ray burst (GRB)
fireballs was anticipated because the highly super-Eddington
luminosities suggest that baryons are expelled. In fact, the baryonic
component that is expected to accompany such an outflow would quench
the $\gamma$-ray emission. This realization led to the popularization
of a model for GRBs in which baryonic kinetic energy was reclaimed at
large radii by internal shocks to be used for making high energy
particles \citep{LE93} and $\gamma$-rays \citep{MR94}. This model
postulated fewer baryons than expected from a-priori estimates of the
super-Eddington flux-driven mass outflow, but it did invoke baryons,
or in any case matter that survived annihilation in the compact
regions closer to the central engine.

The above models should be contrasted with internal shocks models
in which there are few baryons \citep{Eichler94} where the role of
the internal shocks, presumably near the photosphere, is simply to
non-thermalize the spectrum of $\gamma$-rays that dominate the
energy content. In particular, the reclamation of baryonic kinetic
energy by internal shocks for the purposes of making the prompt
$\gamma$-radiation of the GRB itself \citep{MR94} led to the
prediction that there should be a blast wave in the circumburst
medium that would generate afterglow. The discovery of afterglow
appeared to provide enormous support  to the internal shocks model
of \citet{MR94}, who had predicted such an afterglow.

On the other hand, it has never been proven that the $\gamma$-rays and
the ejecta are made by the same components of the outflow.  More
generally, the ratio of $\gamma$-ray to baryonic energy per solid
angle may vary considerably within the outflow in a way that would
have significant consequences both for observations and for theories
of GRB origin. One consequence of GRB baryon anisotropy would be lines
of sight that are more favored than others for GRB detection while
others might be more favorable for seeing early afterglow.  In this
{\it Letter}, we argue in favor of this hypothesis. In \S \ref{BA} we
present arguments in favor of baryon anisotropy. In \S \ref{flat} we
show that the combination of the decaying tail of the prompt emission
and the flat early afterglow light curve from viewing angles slightly
outside the edge of the region within the jet with prominent afterglow
emission can produce the early flat decay phase of the X-ray
afterglows recently observed by {\it Swift}. Our conclusions are
discussed in \S \ref{diss}.

\section{Arguments in Favor of Baryon Anisotropy within the Jet}
\label{BA}

Several arguments have been put forth that the baryon richness of a
GRB fireball relative to the $\gamma$-ray intensity varies with the
viewing angle for a given fireball. \citet{LE93} argued theoretically
that the anatomy of a GRB is likely to be Poynting flux or
$\gamma$-rays emerging from horizon-threading magnetic field lines,
while the baryons might flow out predominantly near the periphery of
the above, along field lines that thread the interface of the inner
accretion disk and the event horizon. On field lines that thread the
accretion disk, which should also have a super-Eddington power output,
there ought to be a ``slow sheath'' of baryons that, while being too
baryon rich to yield a detectable GRB, would yield a contribution to
the afterglow and possibly a ``dirty fireball''. Several observations
have also been interpreted in the context of a dichotomy between the
baryon poor flow lines and the baryon rich ones. Scattering of
$\gamma$-rays from baryon poor regions off more slowly moving baryons
could yield a weak spray of $\gamma$-rays at large angles. Such
scattered emission would be a very small fraction ($\sim 10^{-5}$) of
the total, and would thus be detectable only for very nearby
GRBs. Nevertheless, they would be detectable over a much larger solid
angle, and would be manifested as soft GRBs (no photons well above an
MeV), with smooth light curves from nearby sources. GRB 980425 was a
good example of such a GRB and was interpreted in this light
\citep{Nakamura,EL99}.

Another intriguing result of baryon anisotropy of the GRB fireball
could be that the $\gamma$-ray emission itself - not just the
baryonic content - could have a non-trivial angular profile.  Here
the details have not been worked out yet. It could be, for
example, that most of the emission in the interior of the outflow
is virgin Poynting flux, and that the $\gamma$-rays themselves are
generated preferentially at the periphery of the fireball where
the Poynting flux is somehow tapped by interaction with the slower
baryonic sheath. It is also not yet established under what
conditions virgin Poynting flux creates afterglow.  If a necessary
condition for afterglow is that external protons can execute at
least one gyroradius in the comoving frame within a proper
hydrodynamic timescale, $\sim R/\Gamma c$, then the condition can
be expressed \citep{Eichler03} as $\Gamma \le \Sigma^{1/3}$,
%\begin{equation}
%\Gamma \le \Sigma^{1/3}\ ,
%\end{equation}
where $\Sigma$ is the electric potential energy drop, $e\beta BR$,
across the ejecta in units of $m_p c^2$, and $B$ is the magnetic
field in the lab frame. This would typically imply that $\Gamma
\lesssim 10^4$ for GRBs. However, there are still several
additional considerations that could be relevant.
%enter into the picture.

If the angular structure of the $\gamma$-ray emitting region is more
complicated than a solid cone, then the fraction of the total solid
angle from which the $\gamma$-rays are detectable that corresponds to
off-beam viewing angles increases, thus making such lines of sight
more probable. In this regard, the \citet{Amati02} and
\citet{Ghirlanda04} relations, which correlate the spectral peak
photon energy, $E_{\rm peak}$, and the apparent isotropic equivalent
energy, $E_{\rm\gamma,iso}$, can both be explained as viewing angle
effects \citep{EL04,LE05}. GRBs with $E_{\rm peak}\ll 1\;$MeV are
interpreted as viewed off-beam at angles $\theta>\Gamma^{-1}$ from the
edge of the jet.\footnote{In this case the observed photons are
directed backwards in the comoving frame of the emitting plasma.} This
lowers both the observed $E_{\rm\gamma,iso}$ and $E_{\rm peak}$ in a
way that conforms to the Amati relation \citep{EL04}.  Furthermore,
the value of the jet opening angle that is inferred from the observed
break time in the afterglow and the observed $E_{\rm\gamma,iso}$ is
then slightly overestimated relative to its true value, because the
observed $E_{\rm\gamma,iso}$ is underestimated relative to its true
value. When this slight overestimate is accounted for, the Ghirlanda
relation becomes equivalent to the Amati relation
\citep{LE05}.\footnote{Note that the Amati relation applies only if
{\it most} (or in any case, a fixed fraction) of the total GRB energy
is emitted more or less isotropically in a high Lorentz factor
frame. In cases such as GRB~980425, where only a small fraction,
$f\sim\Omega_o/\Omega_f$, (where $\Omega_o$ and $\Omega_f$ are the
original and final opening angles of the scattered radiation) is
presumed to be scattered by slowly moving (say non-relativistic)
material to large angles, then the observed $E_{\rm\gamma,iso}$ is
lowered by an additional factor f with no further change in $E_{\rm
peak}$, and such sources should be highly distinct outliers to the
Amati relation, as they are observed to be.}

Another recent piece of evidence for baryon anisotropy comes from
the 2004 December 27 giant flare from SGR~1806$-$20. The radio
afterglow \citep{Gaensler05} that followed this event has been
interpreted as being powered by a baryonic mass outflow of $\sim
10^{25}\;$g \citep{Gelfand05} that is probably driven from the neutron
star surface. This tentative conclusion is based on the fact that if
the required mass was, instead, dominated by swept up external medium,
this would require a highly contrived and unrealistic external density
profile in order to explain the observed evolution of the source size,
motion, and flux \citep{GranotSGR05}. Furthermore, this outflow should
have been only mildly relativistic, with $\Gamma\beta\sim 1$, close to
the escape velocity from the surface of the neutron star
\citep{GranotSGR05}. On the other hand, such a large and mildly
relativistic baryonic outflow would have (at least partly) obscured
the prompt $\gamma$-rays from the flare had they been expelled in our
direction. The data can be reconciled, however, by the assumption that
the baryons are ejected in some directions and not others.

\citet{EJ-H05} reconsidered the issue of blast efficiency, making
the assumption that $E_{\rm peak}$ is determined mostly by the
viewing angle effect. The blast efficiency was computed by
previous authors based on the X-ray afterglow at 10 hours. It was
found that when the ratio of blast energy to apparent $\gamma$-ray
energy is tabulated for GRBs with known redshifts, the ratio is of
order unity but with a great deal of scatter.  This scatter,
however, is considerably reduced when the ratio of ``viewing angle
corrected'' $\gamma$-ray energy,\footnote{The form of this
correction assumes that the Ghirlanda relation is due to viewing
angle effects so that the factor $E_{\rm peak}^{3/2}$ is taken to
be a measure of the Doppler factor that corresponds to the viewing
offset angle.} $\propto E_\gamma/E_{\rm peak}^{3/2}$, is compared
with blast wave energy, and the characteristic value of the
$\gamma$-ray to kinetic energy ratio seems to be $\sim 7$. That
is, nearly $90\%$ of the energy goes into radiation
($\gamma$-rays) and only about $10\%-15\%$ goes into the blast
wave.

Even if the $90\%$ gamma ray efficiency was the case for all GRB,
this would raise serious problems for the internal shocks model,
in which the $\gamma$-ray energy is powered by whatever fraction
of the baryonic kinetic energy can be radiated away by the
internal shocks. In particular, it would require the internal
shocks to consistently radiate away nearly $90\%$ of the total
energy within the observed photon energy range and consistently
leave the same small fraction. It is hard to see how internal
shocks, which are by nature erratic, could perform so efficiently
and so consistently. Moreover, even if internal shocks could
consistently convert more than $90\%$ of the kinetic energy into
internal energy, they would still need to put more than $90\%$ of
the internal energy into electrons (and therefore $<10\%$ into
ions) which could radiate this energy within the dynamical (i.e.
expansion) time. Although there in no definitive model for
supercritical ultra-relativistic shocks, this requirement would
run contrary to theoretical expectations. To make matter worse,
Eichler and Jontof-Hutter (2005) found that several outlying GRB
had an even higher gamma ray efficiency $\gtrsim 99\%$. They
suggested that these GRB were viewed along a line of sight that,
even after 10 hours, remained outside the $1/\Gamma$ emission cone
of the baryons.

The idea that many GRBs are viewed slightly off the main intensity
peak by a finite offset angle, $\theta$, was used
\citep{Eichler05} to interpret delayed onset of X-ray afterglow
that was reported by \cite{Piro05}. The delay is simply the time
needed for the flow to decelerate to a Lorentz factor
$\Gamma<1/\theta$, beyond which time the afterglow beam
encompasses our line of sight. This would predict, under the
assumption that the afterglow-generating blast wave and the
$\gamma$-ray beam coincide, that the lower the spectral peak, the
longer it will take for the afterglow to assume its full on-beam
value \citep{Granot02,Granot05}. However, the outliers to the
$\epsilon_b$ - $E_{peak}$ correlation found by Eichler and
Jontof-Hutter (2005) suggest that this correlation could be
contaminated by many GRB that are bright, spectrally hard and have
extensively delayed afterglow onset.

Very recently, \citet{Nousek05} arrived at a similar conclusion
using early X-ray afterglow data taken by {\it Swift}. The early
X-ray afterglows frequently show an intermediate phase ( $t_1 < t
< t_2$) of very flat flux decay at early times (between a few
$10^2\;$s and $\sim 10^4\;$s) that is deficient relative to
expectations from a uniform adiabatic blast wave.  Among this
subset of GRBs, it typically takes the X-ray afterglow a few hours
to attain the values inferred from {\it BeppoSax} data. They note
that for $ t \ll t_2$, the $\gamma -$ray efficiency,
$\epsilon_\gamma$, is typically much higher than previously
inferred \citep{PK01,LRZ04}, yet at $ t \gtrsim t_2$, the
distribution in these efficiencies converges to the narrow range
of values inferred previously.\footnote{Actually, the better
terminology is the blast efficiency,
$\epsilon_b=1-\epsilon_\gamma$. In this class of bursts
$\epsilon_\gamma\sim 90\%$ and the blast efficiencies
$\epsilon_b\sim 10\%$ have greater relative scatter due to their
smaller values.}

Yet another possible effect produced by offset viewing is the spectral
evolution of spikes in the prompt GRB emission, which scale as $E_{\rm
peak}\propto t^{-2/3}$ \citep{Ryde04}. This can be interpreted as the
spike being due to an accelerating blob of matter that scatters
primary radiation into our line of sight. Assuming that the blob is
being accelerated to high $\Gamma$ by an extremely super-Eddington
radiation flux in the direction of the local radiation flux, the
Lorentz factor of the blob scales as $\Gamma\propto R^{1/3}$
\citep{Eichler04}. A spectral peak photon energy which is $E_*$ in the
source frame, is $E_*/\Gamma$ in the blob frame, and $E_{\rm
peak}=E_*/\Gamma^2(1-\beta\cos\theta)$ in the observer's frame. Once
the blob has accelerated to $\Gamma \gg 1/\theta$, then the observed
time scales linearly with radius, $t\propto R$, and $(1-\beta
\cos\theta)$ is asymptotically constant ($\approx\theta^2/2$) so that
$E_{\rm peak}\propto\Gamma^{-2}\propto R^{-2/3} \propto t^{-2/3}$.
The general picture of matter being accelerated by radiation is
consistent with the inference argued above that the matter draws its
kinetic energy from the radiation rather than the other way around.

\section{Delayed Afterglow Onset and Flat Early Light Curves}
\label{flat}

The explanation favored by \citet{Nousek05} for the early flat part of
the X-ray afterglow light curves that were observed by {\it Swift} is
energy injection into the afterglow shock \citep[see
also][]{Zhang05,Panaitescu05}. Here we suggest an alternative
explanation for this early flat phase, namely the flat early afterglow
light curve for viewing angles slightly outside the (rather sharp)
edge of the jet (i.e. outside the regions where the energy per solid
angle in the external shock is large enough to produce bright
afterglow emission).

For such ``off-beam'' lines of sight, the afterglow flux initially
rises, at early times, as the beaming of the radiation away from
the line of sight gradually decreases with time, then rounds off
as the afterglow beaming cone expands enough to include the line
of sight, and finally gradually joins the decaying ``on-beam''
light curve (seen by observers within the jet). For a point
source, the fluxes seen by off-beam and on-beam observers are
related by
\begin{equation}\label{point_source}
F_\nu(\theta,t)\approx a^3F_{\nu/a}(0,at)= a^{3+\beta_X-\alpha_X}F_\nu(0,t)
\end{equation}
 \citep{Granot02}, where $\theta$ is the angle between the source's
velocity and the direction to the observer in the lab frame, $t$ is
the observed time,
$a=(1-\beta)/(1-\beta\cos\theta)\approx(1+\Gamma^2\theta^2)^{-1}$ is
the ratio of the off-beam (at $\theta$) and on-beam ($\theta=0$)
Doppler factors, and the last equality is valid when
$F_\nu(0,t)\propto t^{-\alpha_X}\nu^{-\beta_X}$. Thus, the off-beam
flux is suppressed relative to the on-beam flux by a factor
$\eta=a^{3+\beta_X-\alpha_X}$.  For early times, when $\Gamma\theta\gg
1$, we have $a\approx(\Gamma\theta)^{-2}$ and $\Gamma\propto
t^{-(3-k)/2}$ where $\rho_{\rm ext}\propto R^{-k}$, so that $a\propto
t^{3-k}$, $\eta\propto t^{(3-k)(3+\beta_X-\alpha_X)}$ and
$F_\nu(\theta,t)\propto\eta(t)t^{-\alpha_X}\propto
t^{(3-k)(3+\beta_X)-(4-k)\alpha_X}$. The point source limit is valid
when the angle $\theta$ from the closest point along the edge of the
jet is larger than the typical angular extent of the jet, $\theta_{\rm
jet}$. However, more often the opposite is true,
i.e. $\theta<\theta_{\rm jet}$ (or
$\Gamma_0^{-1}\ll\theta\ll\theta_{\rm jet}$ where $\Gamma_0$ is the
initial Lorentz factor of the jet), in which case the dependence of
$\eta$ on $a$ decreases by one power \citep{EL04,LE05} to
$\eta=a^{2+\beta_X-\alpha_X}$. This implies $\eta\propto
t^{(3-k)(2+\beta_X-\alpha_X)}$ and
$F_\nu(\theta,t)\propto\eta(t)t^{-\alpha_X}\propto
t^{(3-k)(2+\beta_X)-(4-k)\alpha_X}$ at very early times.

\begin{figure}
\centerline{\includegraphics[width=16.0cm]{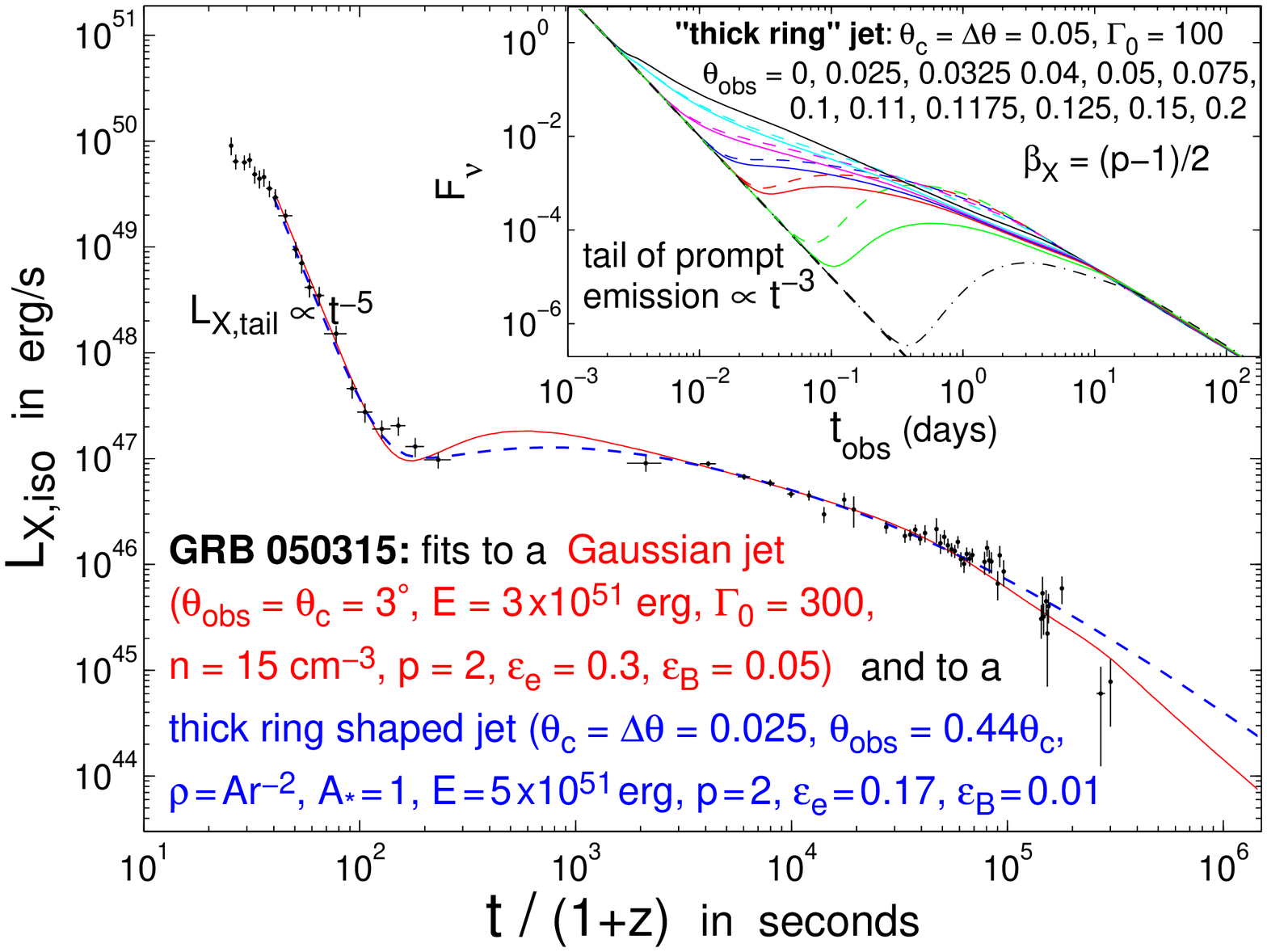}}
%\plotone{f1.eps}
\caption{\label{LC1}A tentative fit to the X-ray light curve of
  GRB~050315 \citep[from][]{Nousek05}, with (i) a Gaussian jet
  \citep[{\it red solid line}, using model 1 of][]{GK03}, and (ii) a
  ring shaped jet, uniform within
  $\theta_c<\theta<\theta_c+\Delta\theta$ \citep[{\it blue dashed
  line}, following the model described in][]{Granot05}. The initial
  fast decay is attributed to the tail of the prompt emission and
  modeled as a power law $\propto t^{-5}$. The inset shows afterglow
  light curves for a ring shaped jet \citep{Granot05}, for different
  viewing angles $\theta_{\rm obs}$ from the jet symmetry axis
  .}
\end{figure}

Fig. \ref{LC1} shows two tentative fits to the X-ray light curve of
GRB~050315, which is perhaps the best monitored X-ray light curve
showing a pronounced early flat phase \citep{Nousek05}. The first fit
({\it red solid line}) is for a Gaussian jet, calculated using model 1
of \citet{GK03}, where lateral expansion is neglected, the jet
dynamics is calculated according to a simplified semi-analytic
formulation of the energy conservation equation, and the light curve
is calculated by integrating over the equal arrival time surface of
photons to the observer, assuming a simple piece-wise power law
synchrotron (and synchrotron self-Compton) spectrum in comoving frame.
The second fit ({\it dashed blue line}) if for a jet with a cross
section in the shape of a thick ring, calculated using the model
developed in \citet{Granot05} where the light curve is calculated by
integrating over the surface of equal arrival time of photons to the
observer, while the jet dynamics are simplified. The latter model was
also used in the {\it inset} of Fig. \ref{LC1}, which shows the light
curves for different viewing angles.  In all cases the early fast
decay is attributed to the tail emission of the prompt GRB
\citep{KP00} and is modeled by a steep power law.  As derived
analytically in the previous paragraph, the light curves for off-beam
viewing angles show a rather sharp rise in the flux at very early
times. However, in most cases the steep early rise of the afterglow
emission is hidden below the steeply decaying tail of the prompt GRB
emission.  Fig. \ref{LC1} demonstrates that the combination of the
decaying tail emission of the prompt GRB and the gently rising or
rounding off afterglow emission from slightly off-beam viewing angles
can produce a flat early phase in the afterglow light curves, similar
to those seen by {\it Swift} in the X-rays \citep{Nousek05}.  In some
cases we expect to even see a gentle rise at very early times.  As can
be seen from Fig. \ref{LC1}, different combinations of jet structure
and viewing angle can generally fit the same observed X-ray light
curve.

\begin{figure}
\centerline{\includegraphics[width=12.5cm]{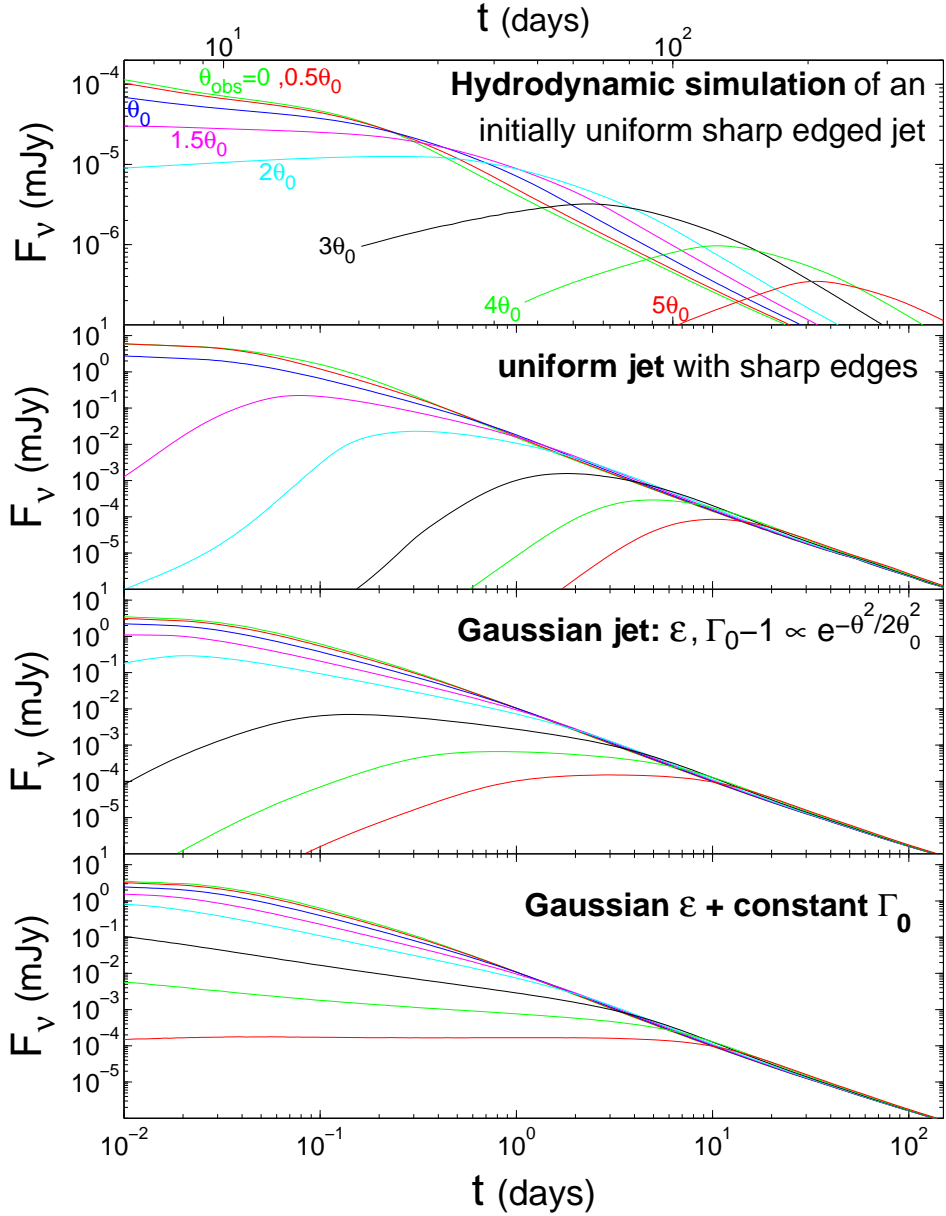}}
%\plotone{f3.eps}
\caption{\label{LC2}Light curves for different jet structures,
  dynamics and viewing angles. The upper panel is from an initially
  uniform jet with sharp edges whose evolution is calculated using a
  hydrodynamic simulation \citep[taken from Fig. 2
  of][]{Granot02}. The remaining three panels are taken from Fig. 5 of
  \citet{GR-RP05}, where a simplified jet dynamics with no lateral
  expansion is used. The middle two panels are for a Gaussian jet, in
  energy per solid angle, and either a Gaussian or a uniform initial
  Lorentz factor, while the bottom panel is for a uniform jet. The
  viewing angles are $\theta_{\rm
  obs}/\theta_0=0,\,0.5,\,1,\,1.5,\,2,\,3,\,4,\,5$ where $\theta_0$ is
  the (initial) half-opening angle for the uniform jet (two upper
  panels) and the core angle for the Gaussian jet (lower two panels).}
\end{figure}

Fig. \ref{LC2} demonstrates the dependence of the early rise in the
afterglow light curves, for off-beam viewing angles, on the jet
structure and dynamics. The three bottom panels are taken from
\citet{GR-RP05} and are calculated using model 1 of \citet{GK03}. The
second panel shows afterglow light curves from a uniform jet with
sharp edges for different viewing angles $\theta_{\rm obs}$ from the
jet symmetry axis. The light curves are similar to those for a wide
ring jet (Fig. \ref{LC1}) for similar off-beam viewing angles.  The
two bottom panels in Fig.  \ref{LC2} are for a Gaussian jet, and show
that the smoother the edges of the jet (both in terms of energy per
solid angle and in terms of the initial Lorentz factor $\Gamma_0$),
the shallower the initial rise in the flux for off-beam viewing
angles.

The upper panel in Fig. \ref{LC2} shows the afterglow light curves for
an initially sharp edged jet whose dynamics were calculated using a
hydrodynamic simulation \citep{Granot01}. The initial conditions were
a cone of half-opening angle $\theta_0$ taken out of the spherical
self-similar solution of \citet{BM76}. The light curves for off-beam
viewing angles, and especially for $1\lesssim\theta_{\rm
obs}/\theta_0\lesssim 2$, are much flatter at early times compared to
those calculated using semi-analytic models for the jet dynamics,
since the shocked external medium at the sides of the jet has a
significantly smaller Lorentz factor than that near the head of the
jet, and therefore its emission is not so strongly beamed away from
off-beam lines of sight. Thus we conclude that very flat (either a
very shallow rise or a very shallow decay) early afterglow light
curves are expected for a realistic jet structure and
dynamics. Combined with the rapidly decaying tail of the prompt GRB
emission, this can nicely reproduce the observed early flat parts of
the X-ray afterglow light curve observed by {\it Swift}.

\section{Discussion}
\label{diss}

We have shown that the early flat phase in the X-ray afterglow
light curves observed by {\it Swift} is broadly consistent with
earlier predictions that afterglow onset might appear delayed to
an ``offset viewer'' - an observer who is outside of the directed
beam of baryons. Many authors ( e.g. Panataiscu and Meszaros,
1998, Panataiscu, Meszros and Rees 1999, Moderski, Sikora and
Bulik 2000, Granot et al 2001, Dalai, Griest and Pruet 2002, Dado,
Dar, and De Rujula 2002) have noted that offset viewing would
suppress early afterglow and presumably prompt gamma rays as well.
\citet{Granot02} predicted that this would be the case for orphan
afterglows, and \citet{GR-RP05} argued that this is expected for
X-ray flashes (and nicely agrees with their pre-{\it Swift}
optical and X-ray early afterglows), assuming that the softening
or non-appearance of prompt $\gamma$-rays in these instances is
due to offset viewing (relative to both the regions of prominent
$\gamma$-ray and afterglow emission, which were assumed to
coincide). \citet{EL04} and \citet{EJ-H05} predicted that this
could also be the case for ``normal'' $\gamma$-ray bursts with a
bright prompt emission if the viewer is in the direct beam of the
gamma rays but not of the baryons. Delays of several minutes in
some afterglows whose onsets were serendipitously caught by the
wide field camera of {\it BeppoSAX} \citep{Piro05} led
\citet{Eichler05} to conjecture that the prompt emission was seen
along an afterglow-inefficient line of sight, and that, if caught
during the first several hours, an even larger fraction of
afterglow onsets would appear delayed. This now seems to be the
case in our view. We have argued here that the stage of flat
decay, often seen within the first few hours of afterglow, can be
attributed to the delayed onset discussed in these earlier papers.

For off-beam viewing angles relative to the region of prominent
afterglow emission, we might expect a correlation between the
flatness of flat decay phase (i.e. its temporal index $\alpha_2$
where $F_\nu\propto t^{-\alpha_2}$) and its luminosity, where a
flatter decay would on average have a lower luminosity. This
indeed appears to be the case judging from Figs. 1 of
\citet{Nousek05}, where the scatter in luminosity is significantly
reduced at late times. This is hard to explain with late time
energy infusion, which would be expected, rather, to increase the
amount of scatter at late times relative to early ones.  If the
same viewing angles are also off-beam relative to the region of
prominent $\gamma$-ray emission, one might expect a correlation
between $\alpha_2$ and the spectral peak photon energy, $E_{\rm
peak}$, or isotropic equivalent energy in $\gamma$-rays,
$E_{\rm\gamma,iso}$. Tables 1 and 2 of \citet{Nousek05} show no
obvious apparent correlation between $\alpha_2$ and
$E_{\rm\gamma,iso}$. The lack of such a correlation would suggest,
under the off-beam interpretation, that {\it the regions of
prominent afterglow emission and of prominent $\gamma$-ray
emission do not coincide}, as we have suggested in this work.
Furthermore, some of the GRBs recorded by {\it Swift}, such as GRB
050315, have a rather large $E_{\rm\gamma,iso}$ {\it as well as}
long stages of very flat decay, and could be interpreted in our
view as being due to lines of sight along which the afterglow
emission was intrinsically weak. Such a weak afterglow emission
can easily be attributed to a paucity of baryons relative to
$\gamma$-rays in the outflow along our line of sight.

\acknowledgements The authors gratefully acknowledge a Center of
Excellence grant from the Israel Science Foundation, a grant from
the Israel-U.S. Binational Science Foundation and support from the
Arnow Chair of Theoretical Astrophysics. This research was
supported by the US Department of Energy under contract number
DE-AC03-76SF00515 (J.~G.).

\newpage

\newpage

\end{document}